\documentclass[twoside]{article}
\usepackage{PRIMEarxiv}

\usepackage[utf8]{inputenc} % allow utf-8 input
\usepackage[T1]{fontenc}    % use 8-bit T1 fonts
\usepackage{hyperref}       % hyperlinks
\usepackage{url}            % simple URL typesetting
\usepackage{booktabs, multicol, multirow}       % professional-quality tables
\usepackage{amsfonts, amsmath, amssymb}       % blackboard math symbols
\usepackage{nicefrac}       % compact symbols for 1/2, etc.
\usepackage{microtype}      % microtypography
\usepackage{lipsum}
\usepackage{fancyhdr}       % header
\usepackage{graphicx}       % graphics
\graphicspath{{media/}}     % organize your images and other figures under media/ folder
\usepackage{setspace}
\usepackage[round]{natbib}
\usepackage{xcolor}
\usepackage{lineno}

\newtheorem{theorem}{Theorem}

\newtheorem{assumption}{Assumption}
\newtheorem{definition}{Definition}
\newtheorem{proof}{Proof}

\newtheorem{remark}{Remark}
\title{A Framework for Covariate-Adjusted Bivariate Causal Discovery
}

\author{
  Soumik Purkayastha \\
  Department of Biostatistics and Health Data Science, \\
  University of Pittsburgh, \\
  Pittsburgh, USA.\\ 
  \texttt{soumik@pitt.edu} \\
   \And
  Peter X.-K. Song \\
  Department of Biostatistics, \\
  University of Michigan, \\
  Ann Arbor, USA\\
  \texttt{pxsong@umich.edu} \\
}

\begin{document}
\doublespacing
%\linenumbers
\maketitle

\begin{abstract}
Ascertaining causal direction from observational data is a fundamental challenge in scientific inquiry. Of particular interest is the problem of covariate-adjusted bivariate causal discovery, i.e., determining the causal direction between $X$ and $Y$ in the presence of $\mathbf{Z}$. While unadjusted bivariate causal discovery has seen significant advances \citep{hoyer2008, ni2022bivariate}, there is a lack of methodology dealing with real-world bivariate relationships, which are often modulated by a set of covariate(s), $\mathbf{Z}$. Building on previous work in \cite{purkayastha2025}, we introduce a novel, nonparametric framework for the covariate-adjusted bivariate causal discovery problem and propose a conditional asymmetry coefficient to track said direction of causation
We develop a robust estimation procedure using kernel-based conditional density estimation with cross-fitting and also provide rigorous uncertainty quantification using a nonparametric kernel smoothing technique, addressing a key limitation of many existing algorithms. As a key application, we demonstrate how this framework can be applied to the problem of collider detection, a persistent challenge in causal structure learning. Simulation studies show our method's superior performance in identifying causal structures. We apply our approach to an epigenetic study \citep{Perng_2019}, investigating the role of blood pressure in regulating the effects of \emph{DNA} methylation. In summary, our work offers methodological advancement by providing a robust, inferential toolkit for dissecting complex, moderated bivariate causal relationships in observational data.
\end{abstract} 
\keywords{Asymmetry \and Causal discovery \and Collider detection \and Conditional density estimation \and Information theory \and Nonparametric kernel smoothing}

\section{Introduction}

Inferring causal directionality from observational data is a central goal across numerous disciplines, including biomedical studies, economics, and epidemiology \citep{spirtes_2016, glymour2019review}. While significant progress has been made in the bivariate case of determining whether $X \rightarrow Y$ or $Y \rightarrow X$ \citep{choi2020, tagasovska2020, ni2022bivariate}, these methods often fall short in practice where bivariate relationships are rarely isolated, i.e., when the causal link between two variables, $X$ and $Y$, is often moderated or influenced by a third variable, $\mathbf{Z}$, denoted by $X \leftarrow Y \mid \mathbf{Z}$, or $X \rightarrow Y \mid \mathbf{Z}$. This introduces a more complex and realistic problem, namely, covariate-adjusted bivariate causal discovery.

The challenge lies in distinguishing the true causal direction when the mechanism itself may depend on $\mathbf{Z}$. This problem extends foundational work on Additive Noise Models (ANM) of the form $Y = g(X) + \epsilon$, which posit an asymmetry in the independence of the noise term $\epsilon$ relative to the cause $X$ \citep{hoyer2008, zhang2012kernelbasedconditionalindependencetest}. ANMs fail in the context of deterministic systems denoted by $Y = g(X)$, which gave rise to the seminal framework of Information Geometric Causal Inference (IGCI) \cite{daniusis2012inferring, janzing2012information}. However, the IGCI framework examines directionality for the relationship $Y = g(X)$ for bijective $g: [0, 1] \rightarrow [0, 1]$ alone. In \cite{purkayastha2025}, we establish a self-contained theoretical framework of Generative Exposure Mappings (GEMs) that significantly broadens the existing IGCI framework. Our GEM framework broadens the scope of the IGCI framework by considering a much richer class of generating functions (GFs) that includes bijective functions. In this paper we provide an extension of the GEM framework that makes several unique contributions to address the limitations of existing methods.
\begin{enumerate}
    \item Our approach moves beyond the constraints of additive noise models by directly formulating the effect as a general, non-linear function of the cause, conditional on the covariates, i.e., $Y=g(X, \mathbf{Z} = \mathbf{z})$ for $\mathbf{z} \in \mathcal{Z}$, where $\mathcal{Z}$ is the support of $\mathbf{Z}$. This provides greater flexibility to capture complex relationships potentially moderated by $\mathbf{Z}$. 
    \item We develop a complete inferential pipeline that provides confidence intervals and formal hypothesis tests for the existence of causal pathways, allowing for principled uncertainty quantification.  
\end{enumerate} We present methodology for both estimation and inference of covariate-adjusted bivariate causal discovery. We employ a kernel-based conditional density estimator \citep{hyndman_estimating_1996} with a data-splitting and cross-fitting procedure \citep{chernozhukov2018double} for robust estimation. We then use a nonparametric kernel smoothing technique \citep{cleveland1988locally, fan2018local} to construct confidence intervals and perform hypothesis tests on the proposed metric that tracks directionality. This provides a rigorous, data-driven approach to uncovering moderated bivariate causal relationships in complex observational data structures.

A crucial application of this general framework is in identifying collider structures, given by $X \rightarrow \text{COL} \leftarrow Y$, where a variable $\text{COL}$ is the collider between $X$ and $Y$. Colliders have been studied in various fields including biomedical studies \citep{holmberg2022collider, Weiskopf2023}, economics \citep{Schneider2020}, and epidemiology \citep{Banack2023}. In this paper, we restrict our attention to v-structures, where there are causal pathways from both $X$ to $\text{COL}$ and from $Y$ to $\text{COL}$, but no direct pathway between $X$ and $Y$. Adjusting for a collider famously induces spurious associations between its parents, making its correct identification vital for valid statistical modeling \citep{pearl2009}. Here, we re-frame collider detection as a special case of our more general methodology. By treating one parent as the `cause' and the other as the `covariate', we check for the pathway $X \rightarrow \text{COL}$ conditional on $Y$, and vice-versa. Confirming both pathways provides strong evidence for the collider structure. Prior attempts at collider structure learning have resulted in two broad classes of algorithms that estimate some underlying causal structure. The two classes of algorithms are (i) constraint-based structure learning algorithms which learn a set of causal structure(s) that satisfy the conditional independences embedded in the data \citep{spirtes2000causation, margaritis2003learning}, or (ii) score-based structure learning algorithms that learn causal structures by  maximizing a scoring criterion, with low scores assigned to structures that embed incorrect conditional independences in the data \citep{cohen1994probabilistic}. Additionally, there are some  hybrid algorithms that exploit principled ways to combine constraint-based and score-based algorithms \citep{tsamardinos2006max, friedman2013learning}. %To some extent, these approaches are sensitive to latent confounding. 
These methods do not provide statistical inference on the estimated pathways. 
Our methodology is an attempt at  detecting collider structures in non-linear generative models that provides uncertainty quantification.

The rest of this paper is organized as follows. In Section \ref{sec:prelims} we consider mappings of the form $Y = g(X, \mathbf{Z})$ and introduce the notion of Conditional GEM (CGEM) between exposure $X$ and outcome $Y$, moderated by $\mathbf{Z}$. In Section \ref{sec:strong}, we define the conditional asymmetry coefficients, denoted by $C_{X \rightarrow Y \mid \mathbf{Z}}$ and present theoretical support in favour of the coefficient when evaluating directionality between $(X, Y)$ in the presence of $\mathbf{Z}$. Section \ref{sec:noise_tol} describes how our proposed methodology works not just for deterministic systems but is also robust to noise contamination upto a certain threshold. Section \ref{sec:estim_inf} presents estimation and inference details; we implement a kernel conditional density estimation technique to estimate $C_{X \rightarrow Y \mid \mathbf{Z}}$, followed by a nonparametric kernel smoothing technique that allows us to perform inference on the estimated pathways. Section \ref{sec:col_app} details an application of our proposed methodology to collider detection. Section \ref{sec:sim} and Section \ref{sec:rda} illustrate the performance of the proposed framework through empirical studies and a real data application respectively. To conclude, we discuss key findings, outline current limitations, and propose potential future work in Section \ref{sec:conclusion}.

\section{Formulation} \label{sec:prelims}

\subsection{The conditional generative exposure mapping setup}

Let us consider a population where for each unit, we observe a triplet of continuous variables $(X,Y, \mathbf{Z})$. The distribution of the cause $X$, conditional on  any given $\mathbf{Z} = \mathbf{z}$ is governed by its conditional probability density function (PDF), $f(x \mid \mathbf{z})$. The generative process is then specified by a CGEM; for the causal direction $X\rightarrow Y \mid \mathbf{Z}$, this takes the form
\begin{equation} \label{eq:cgem}
    Y=g(X,\mathbf{Z} = \mathbf{z})=g_{\mathbf{z}}(X), \mathbf{z} \in \mathcal{Z}. 
\end{equation} In this CGEM, the behavior of the outcome $Y$ for a given $\mathbf{z}$ is dictated entirely by the conditional distribution of the cause, $f(x \mid \mathbf{z})$, and the specific generative function, $g_{\mathbf{z}}$. The primary challenge is to infer the correct causal direction, i.e., whether $X$ causes $Y$ conditional on $\mathbf{Z}$ or vice-versa—from observational data. Our framework addresses this by leveraging information-theoretic notions of conditional entropy to quantify the asymmetry induced by the CGEM. 

\subsection{Basic Information Theoretic Concepts}
In preparation for the exposition of conditional generative exposure models, we introduce some important information theoretic concepts. 
For the generative exposure mapping $Y = g(X, \mathbf{Z})$, let the continuous random variables $X, Y, \mathbf{Z}$ have marginal densities $f_X, f_Y, f_{\mathbf{Z}}$ respectively. Further, let $\mathcal{X}, \mathcal{Y}$, and $\mathcal{X}$ denote the respective compact support sets of $X, Y$, and $\mathbf{Z}$. The {conditional entropy functions} of $X\mid \mathbf{Z} = \mathbf{z}$ and $Y \mid \mathbf{Z} = \mathbf{z}$ for a specified value $\mathbf{z} \in \mathcal{Z}$ are given as follows:
\begin{equation}
    \begin{aligned}\label{eq:cef_def}
    H(X \mid \mathbf{Z} = \mathbf{z}) = H(X \mid \mathbf{z}) &:= \int_{x\in \mathcal{X}_\mathbf{z}}{ - \log \left(  {f(x \mid \mathbf{z})} \right) } f(x \mid \mathbf{z}) dx,\\
     H(Y \mid \mathbf{Z} = \mathbf{z}) = H(Y \mid \mathbf{z}) &:= \int_{y\in \mathcal{Y}_\mathbf{z}}{ - \log \left(  {f(y \mid \mathbf{z})} \right) } f(y \mid \mathbf{z}) dy,
\end{aligned}
\end{equation}
where $f(x \mid \mathbf{z})$ and $f(y \mid \mathbf{z})$ denotes the conditional densities of $X \mid \mathbf{Z} = \mathbf{z}$ and $Y \mid \mathbf{Z} = \mathbf{z}$ having support $\mathcal{X}_\mathbf{z}$ and $\mathcal{Y}_\mathbf{z}$ respectively. These functions measure randomness in the conditional distribution of \(X\) and \(Y\) respectively for a given value of \(\mathbf{Z}= \mathbf{z}\). 

\section{Asymmetry in CGEMs}
\label{sec:strong}
In this section we derive a legitimate population-level measure that will be used to confirm ordering or asymmetry under a hypothesized CGEM. Let us begin with the case of a CGEM as described by \eqref{eq:cgem}; we will consider a specific classification of the generating function (GF), denoted by $\mathcal{G}$, as defined below. 
\begin{definition} \label{def:contract}
  Let $\mathcal{G}$ be a class of monotonic and differentiable functions on a compact set $\gamma \subset \mathbb{R}$. For any $g \in \mathcal{G}$, let the following integral be well-defined and satisfy the inequality
  \begin{equation*}
      \exp \left({|\gamma|}^{-1} \int_\gamma \log |\nabla g(x)|\, d x\right)<1, 
  \end{equation*}
  where $|\gamma|$ denotes the Lebesgue measure of $\gamma$, and $\nabla$ denotes the gradient of $g$. We interpret any function $g \in \mathcal{G}$ to exhibit \emph{contracting dynamics} over the support $\gamma$, meaning that the {geometric mean} of $|\nabla g(x)|$ over $\gamma$ is strictly less than one.
\end{definition}

\begin{definition} \label{rem:dynamics}
    Alternatively, we will consider a class of monotonic functions $\mathcal{G}^\star$ with `expanding dynamics' over support $\gamma$; for any $g^\star \in \mathcal{G}^\star$, we have
    \begin{equation*}
      \exp \left( {|\gamma|}^{-1} \int_\gamma \log |\nabla g^\star(x)| d x\right)>1, 
  \end{equation*} i.e., the {geometric mean} of $|\nabla g(x)|$ over $\gamma$ is strictly more than one. 
\end{definition}

An identifiability assumption required to unearth the induced asymmetry from a CGEM is that the conditional distribution of exposure $X \mid \mathbf{Z} = \mathbf{z}$  and the mechanism of the GF $g_{\mathbf{z}}$  do not influence each other when yielding the outcome $Y \mid \mathbf{Z} = \mathbf{z}$, which we quantify through functional orthogonality, as defined below. In the literature of functional analysis, orthogonality is appropriate to characterize the notion of ``no influence'', which will be adopted in this paper. Similar assumptions are found in \cite{Janzing2008, daniusis2012inferring, mooij_2016, purkayastha2025}. 
\begin{assumption}
\label{ass:Iden1} 
Let $g_{\mathbf{z}}$ be a GF in \eqref{eq:cgem} and $f(x \mid \mathbf{z})$ be the conditional density that governs the input $X \mid \mathbf{Z} = \mathbf{z}$ with support $\mathcal{X}_\mathbf{z}$ for any $\mathbf{z} \in \mathcal{Z}$. We assume that the following condition is satisfied 
\begin{equation*}
    \int_{\mathcal{X}_\mathbf{z}}  \log \left(\lvert \nabla g_{\mathbf{z}}(x) \rvert \right)f(x \mid \mathbf{z})dx =  \lvert\mathcal{X}_\mathbf{z} \rvert ^{-1} \int_{\mathcal{X}_\mathbf{z}} \log \left(\lvert \nabla g_{\mathbf{z}}(x) \rvert \right)dx,
\end{equation*}
where the $\nabla$ operator denotes the gradient of $g_{\mathbf{z}}$ with respect to its argument.
\end{assumption}
\begin{remark}
\label{rem:ass1_satisfy}
Assumption \ref{ass:Iden1} automatically holds when $X \mid \mathbf{Z} = \mathbf{z} \sim \mathcal{U}(0,1)$. This is analogous to the assumption of randomization or no confounding in the school of Neyman-Rubin causality \citep{rubin2005causal}; the uniform distribution on $X \mid \mathbf{Z} = \mathbf{z}$ is analogous to randomization in an experiment that leads to no bias in operating exposure $X \mid \mathbf{Z} = \mathbf{z}$. Assumption~\ref{ass:Iden1} may also hold under non-uniform distributions such as Beta$(a,b)$, provided there is a functional balance between the shape of $f(X \mid \mathbf{z})$ and the transformation $g_{\mathbf{z}}$. See \cite{purkayastha2025} for more details. 
\end{remark}

\begin{remark}
To interpret Assumption \ref{ass:Iden1}, consider the quantities $\log \left(\lvert \nabla g_{\mathbf{z}}(X) \rvert \right)$ and $f(X \mid \mathbf{Z} = \mathbf{z})$ as random variables on support $\mathcal{X}_{\mathbf{z}}$. Then, their covariance with respect to the uniform distribution on $\mathcal{X}_{ \mathbf{z}}$ is proportional to 
\begin{equation*}
    \int_{\mathcal{X}_{\mathbf{z}}}  \log \left(\lvert \nabla g_{\mathbf{z}}(x) \rvert \right)f(x \mid \mathbf{Z} = \mathbf{z})dx - \lvert\mathcal{\mathcal{X}_{\mathbf{z}}}\rvert ^{-1} \int_{\mathcal{X}_{\mathbf{z}}} \log \left(\lvert \nabla g_{\mathbf{z}}(x) \rvert \right)dx \int_{\mathcal{X}_{\mathbf{z}}}  f(x \mid \mathbf{Z} = \mathbf{z}) dx, 
\end{equation*} where the last integral is equal to one. Consequently, the equality in Assumption \ref{ass:Iden1} expresses that the density of exposure, given by $f(x \mid \mathbf{Z} = \mathbf{z}) $, and the (log-transformed) dynamics of the generative mechanism, given by $\nabla g_{\mathbf{z}}(x)$, are uncorrelated mechanisms of nature, linking with the notion of algorithmic independence \citep{Janzing2008}.
\end{remark}

We build on the principle of orthogonality as described by Assumption \ref{ass:Iden1}, which posits that the distribution of a cause is independent of the mechanism that generates the effect. In our CGEM framework, this implies that for the true causal direction, the conditional distribution of the cause, $f(x \mid \mathbf{z})$, and the dynamics of the generating mechanism, $g_{\mathbf{z}}$, are uncorrelated. This leads to a testable asymmetry in the conditional entropies of $X \mid \mathbf{Z} = \mathbf{z}$ and $Y \mid \mathbf{Z} = \mathbf{z}$. For $g_{\mathbf{z}} \in \mathcal{G}$ (or $g_{\mathbf{z}} \in \mathcal{G}^\star$), the change of variables formula for differential entropy gives rise to the following equality. 
\begin{equation*}
    H(Y \mid \mathbf{Z} = \mathbf{z})=H(X \mid \mathbf{Z} = \mathbf{z})+\mathbb{E}_{X \mid \mathbf{Z} = \mathbf{z}}\left[\log \left(\left|\nabla g_{\mathbf{z}}(X)\right|\right)\right].
\end{equation*}
Under the independence principle specified by Assumption \ref{ass:Iden1}, the expectation term simplifies to the integral in our definitions of contracting or expanding dynamics, making it strictly negative or positive. This motivates the development of our key measure, the conditional asymmetry coefficient. In \cite{purkayastha2025} we propose an asymmetry coefficient to investigate directionality in bivariate $(X, Y)$; motivated by this nomenclature, we now define the conditional asymmetry coefficient as follows. 
\begin{definition} The conditional asymmetry coefficients are defined as:
\begin{equation}
        \begin{aligned} 
        C_{X \rightarrow Y \mid \mathbf{Z}} &:=\min _{\mathbf{z} \in \mathcal{Z}} \ H(X \mid \mathbf{z}) - H(Y \mid \mathbf{z})\\
        C_{X \rightarrow Y \mid \mathbf{Z}}^{\star} &:=\max _{\mathbf{z} \in \mathcal{Z}} \ H(X \mid \mathbf{z}) - H(Y \mid \mathbf{z}). 
\end{aligned}
    \end{equation}
\end{definition}
If Assumption \ref{ass:Iden1} holds for all $\mathbf{z} \in \mathcal{Z}$, a positive value of $C_{X \rightarrow Y \mid \mathbf{Z}}$ for GFs with contracting dynamics, or a negative value of $C^{\star}_{X \rightarrow Y \mid \mathbf{Z}}$ for GFs with expanding dynamics provides evidence for the causal direction $X \rightarrow Y \mid \mathbf{Z}$. We formalize this through the following hypothesis to confirm or negate the presence of the pathway $X \rightarrow Y \mid \mathbf{Z}$ for conditional generative mappings with contracting dynamics: 
\begin{equation} \label{eq:hypo_cont}
\begin{aligned}
    H_{0}:  C_{X \rightarrow Y \mid \mathbf{Z}} \leq 0 \quad &\text{ vs }\quad  H_{1}: C_{X \rightarrow Y \mid \mathbf{Z}} > 0.
\end{aligned}
\end{equation} Analogously, the following hypothesis is used to confirm or negate the presence of the pathway $X \rightarrow Y \mid \mathbf{Z}$ conditional generative mappings with expanding dynamics
\begin{equation} \label{eq:hypo_exp}
\begin{aligned}
    H_{0}:  C^\star_{X \rightarrow Y \mid \mathbf{Z}} \geq 0 \quad &\text{ vs }\quad  H_{1}: C^\star_{X \rightarrow Y \mid \mathbf{Z}} < 0.
\end{aligned}
\end{equation}

\section{Robustness of CGEMs to noise contamination} \label{sec:noise_tol}
Real-world data is inevitably noisy. We extend our framework to the noise-perturbed CGEM, where the observed outcome $Y^\star$ is a contaminated version of the true outcome $Y=g_{\mathbf{z}}(X)$, 
\begin{equation*}
    Y^*=Y+\sqrt{\sigma} \epsilon=g_{\mathbf{z}}(X)+\sqrt{\sigma} \epsilon,
\end{equation*} where $\epsilon \sim N(0, 1)$ and is independent of $(X,\mathbf{Z})$. Our framework's validity depends on whether the entropy ordering is preserved despite this noise.

\subsection{Contamination for GFs with Contracting Dynamics}
For GFs with contracting dynamics ($g \in \mathcal{G}$), we require $H(Y^\star \mid \mathbf{z})<H(X \mid \mathbf{z})$. A modified version of de Bruijn's identity for conditional densities states that for a fixed $\mathbf{z}$, we have the following. 
\begin{equation*}
    \frac{\partial H(Y + \sqrt{t}\epsilon \mid \mathbf{Z}=\mathbf{z})}{\partial t} = \frac{I(Y + \sqrt{t}\epsilon \mid \mathbf{Z}=\mathbf{z})}{2}, 
\end{equation*} where $I(\cdot|\mathbf{z})$ is the conditional Fisher information. Integrating from 0 to $\sigma$ and applying the convolution inequality for Fisher information yields an upper bound on the entropy increase due to noise, given as follows.
\begin{equation*}
    H(Y^*|\mathbf{z}) - H(Y|\mathbf{z}) \le \frac{1}{2}\log(\sigma I(Y|\mathbf{z}) + 1). 
\end{equation*} To ensure the correct causal direction is identified, the entropy of the contaminated outcome must not exceed the entropy of the cause, i.e., $H(Y^\star|\mathbf{z}) < H(X|\mathbf{z})$. This imposes an upper bound on the tolerable noise variance $\sigma < \min_{\mathbf{z} \in \mathcal{Z}} \frac{\exp(2 C_{X \to Y \mid \mathbf{z}}) - 1}{I(Y|\mathbf{z})}$. The numerator represents the "signal" of the causal asymmetry. This result provides a formal robustness guarantee of our method being effective in low-noise regimes where the noise variance is bounded by this signal-to-noise ratio for $\mathbf{z} \in \mathcal{Z}$.

\subsection{Contamination for GFs with Expanding Dynamics}

For GFs with expanding dynamics ($g \in \mathcal{G}^\star$), the argument is more direct. Here, we know that $H(Y \mid \mathbf{z})>H(X \mid \mathbf{z})$. Since $\epsilon$ is independent of $X$ and $Z$, the entropy power inequality \citep{Cover_2005} states that adding noise can only increase the conditional entropy, i.e., 
\begin{equation*}
    H(Y^*|\mathbf{z}) =  H(Y + \sqrt{\sigma} \epsilon \mid \mathbf{z}) \geq H(Y  \mid \mathbf{z}) > H(X \mid \mathbf{z}). 
\end{equation*} In other words, since the entropy of the noise-free outcome already exceeds that of the cause, the entropy of the contaminated outcome will as well. Therefore, for mechanisms with expanding dynamics, our framework is robust to the addition of independent noise without requiring a specific signal-to-noise ratio.

\section{Estimation and inference} \label{sec:estim_inf}

In this section we describe how we use a conditional density estimation technique \citep{hyndman_estimating_1996, bashtannyk_bandwidth_2001, hyndman2002nonparametric} and a local polynomial modeling and smoothing technique \citep{cleveland1988locally, fan2018local} to estimate and perform inference on ${C}_{X \rightarrow Y \mid \mathbf{Z}}$. 

\subsection{Kernel Estimation of Conditional Densities}\label{subsec:kcde}
In this section, we review a technique to estimate the conditional density of $X \mid \mathbf{Z} = \mathbf{z}$ for a  sample given by $\{(X_i, \mathbf{Z}_i), 1 \leq i \leq n\}$ that are independent and identically distributed with joint density $f_{X\mathbf{Z}}$, whose marginal densities are denoted by $f_X$ and $f_\mathbf{Z}$, with support $\mathcal{X}$ and $\mathcal{Z}$ respectively. We will use this estimation technique repeatedly to estimate the conditional asymmetry coefficients $C_{X \rightarrow Y \mid \mathbf{Z}}$ using the sample $\{(X_i, Y_i, \mathbf{Z}_i), 1 \leq i \leq n\}$. 

For ease of exposition in this section, we will suppress the subscript and let $f$ denote the marginal density of $\mathbf{Z}$. Let \(f(x \mid \mathbf{z})\) be the conditional density of \(X\) given \(\mathbf{Z} = \mathbf{z}\). We utilize the kernel conditional density estimator proposed and studied by \cite{hyndman_estimating_1996, bashtannyk_bandwidth_2001, hyndman2002nonparametric}, which is based on locally fitting a log-linear model to always produce a non-negative density estimator. An algorithm  for selecting a bandwidth for the kernel conditional density estimator was proposed and implemented in \cite{hdrcde}. We are interested in applying this method to estimate \(f(x \mid \mathbf{z})\). Below we present a brief overview of the estimation methodology. 

Let \(K\) be a symmetric density function on \(\mathbb{R}\) and \(K_b(u) = b^{-1} K(u/b)\) with bandwidth $b > 0$. Note that as \(b \to 0\), we have $\mathbb{E}\{K_b(X_i - x) \mid  \mathbf{z}\} - f(X \mid \mathbf{z}) = O(b^2)$, which suggests that \(f(x \mid \mathbf{z})\) can be estimated consistently by regressing \(K_b(X_i - x)\) on \(\mathbf{Z}_i - \mathbf{z}\). The estimation technique proposed by \cite{hyndman_estimating_1996} is carried out by minimizing the following kernel weighted loss function: 
\begin{equation*}
    R(\theta; \mathbf{z}, x; b, h) := \sum_{i=1}^n \left\{ K_b(X_i - x) - A(\mathbf{Z}_i - \mathbf{z}, \mathbf{\theta}) \right\}^2 W_h(\mathbf{Z}_i - \mathbf{z}), 
\end{equation*}
where $A(\mathbf{z}, \mathbf{\theta}) = \ell(\theta_0 + \theta_1^\prime \mathbf{z})$. The kernel weighting function is given by \(W_h(u) = h^{-1} W(\mathbf{u}/h)\) with bandwidth \(h > 0\). Using \(\ell(u) = \exp(u)\) is the popular choice, as suggested by \cite{hyndman2002nonparametric}. Note that there are two smoothing parameters: \(h\) controls the smoothness between conditional densities in the \(\mathbf{z}\) direction (the smoothing parameter for the regression) and \(b\) controls the smoothness of each conditional density in the \(x\) direction. We set $K$ and $W$ to be Gaussian kernels and follow the bandwidth selection algorithm presented by \cite{hyndman2002nonparametric} to compute bandwidths $\hat{b}$ and $\hat{h}$, which are then plugged in to obtain the kernel conditional density estimator $\hat{f}(x \mid \mathbf{z})$, given by:  
\begin{equation} \label{eq:cde_def}
    \hat{f}(x \mid \mathbf{z}) := A(0, \hat{\theta}_{x\mathbf{z}}) = \ell(\hat{\theta}_0), \text{ where }  \hat{\theta}_{x\mathbf{z}} :=  \operatorname{argmin}_{\theta \in {\Theta}} R(\theta; \mathbf{z}, x; \hat{b}, \hat{h}), 
\end{equation} where $\Theta$ is the parameter space and $(x, \mathbf{z}) \in \mathcal{X} \times \mathcal{Z}$. The resulting estimator defined in (\ref{eq:cde_def}) is scaled to ensure its integration is equal to 1. Hence, we use the kernel-based conditional density estimator proposed in \cite{hyndman2002nonparametric} to estimate the conditional density, which is subsequently used to obtain a consistent estimator $\hat{H}(X \mid \mathbf{z})$ of the conditional entropy function $H(X \mid \mathbf{z})$ using a sample splitting and cross-fitting technique.

\subsection{Split-and-fit: Estimation of the conditional entropy function}
\label{subsec:consistent_estim}
We invoke a data splitting and cross-fitting technique to estimate the relevant entropy terms. We first split the available data $\mathcal{D} := \left\{(X_1, \mathbf{Z}_1), \ldots, (X_n, \mathbf{Z}_{n}) \right\}$ into two equal-sized, disjoint sets denoted by $\mathcal{D}_1$ and $\mathcal{D}_2$. For notational convenience, we assume the sample size $n$ is even. Using the data split $\mathcal{D}_1$, we obtain an estimate of the conditional density function $\hat{f}_{1}(x | \mathbf{z})$ using the method described in Section \ref{subsec:kcde}. Interchanging the roles of the data splits, we use $\mathcal{D}_2$ to obtain a second estimated conditional density, $\hat{f}_{2}(x | \mathbf{z})$. The estimated density functions $\hat{f}_1$ and $\hat{f}_2$ are then evaluated on data from the opposite split ($\mathcal{D}_2$ and $\mathcal{D}_1$, respectively) to obtain two estimates of the conditional entropy function.
\begin{equation*}
    \begin{aligned}
    \hat{H}_1(X \mid \mathbf{Z} = \mathbf{z}) &= - \frac{1}{n/2}\sum_{i \in \mathcal{D}_1} \ln \left(\hat{f}_{2} \left( {X}_{i} \mid \mathbf{z} \right) \right), \\
    \hat{H}_2(X \mid \mathbf{Z} = \mathbf{z}) &= - \frac{1}{n/2}\sum_{i \in \mathcal{D}_2} \ln \left(\hat{f}_{1} \left( {X}_{i} \mid \mathbf{z} \right) \right).
    \end{aligned}
\end{equation*}
Averaging these two estimates gives the final cross-fitted estimate of the conditional entropy function:
\begin{equation}\label{eq:cef_estim}
\begin{aligned}
    \hat{H}(X \mid \mathbf{Z} = \mathbf{z}) &= \frac{1}{2}\left\{\hat{H}_1(X \mid \mathbf{Z} = \mathbf{z}) + \hat{H}_2(X \mid \mathbf{Z} = \mathbf{z}) \right\} \\
    &= -\frac{1}{n} \left\{ \sum_{i \in \mathcal{D}_1} \ln \left(\hat{f}_{2} \left( {X}_{i} \mid \mathbf{z} \right) \right) + \sum_{i \in \mathcal{D}_2} \ln \left(\hat{f}_{1} \left( {X}_{i} \mid \mathbf{z} \right) \right) \right\}.
\end{aligned}
\end{equation} One can similarly obtain $\hat{H}(Y \mid \mathbf{Z} = \mathbf{z})$, and subsequently, $\hat{C}_{X \rightarrow Y \mid \mathbf{Z} = \mathbf{z}}$ for $\mathbf{z} \in \mathcal{Z}$. The following theorem establishes the consistency of this estimator. We present the necessary technical assumptions in Appendix \ref{app:app2} to support Theorem \ref{thm:cef_estim_consistency}.

\begin{theorem} \label{thm:cef_estim_consistency}
    Suppose assumptions \ref{assum2} - \ref{assum6} hold. Then, for any $\mathbf{z}$ in the support of $\mathbf{Z}$, the estimator $\hat{H}(X \mid \mathbf{z})$ given by (\ref{eq:cef_estim}) is consistent, i.e., $\hat{H}(X \mid \mathbf{z}) \overset{a.s.}{\rightarrow} {H}(X \mid \mathbf{z})$ as $n \rightarrow \infty$.
\end{theorem}
See Appendix \ref{app:app2} for the proof.

\subsection{Grid-based evaluation of the conditional asymmetry coefficient}

Using the cross-fitting routine outlined in Section \ref{subsec:consistent_estim}, we can obtain consistent pointwise estimates of the conditional entropy, $\hat{H}(X \mid \mathbf{Z} = \mathbf{z})$, for any given conditioning point $\mathbf{z}$ in its support $\mathcal{Z}$. To estimate, and perform inference on $C_{X \rightarrow Y \mid \mathbf{Z}}$, we first define an evaluation grid of points spanning the support of $\mathbf{Z}$. Let $\mathcal{G}_\mathbf{Z}$ be a grid of $n_e$ points within the support of $\mathbf{Z}$:
\begin{equation*}
    \mathcal{G}_\mathbf{Z} := \left\{\mathbf{z}^e_{j} \in \mathcal{Z}, 1 \leq j \leq n_e \right\}.
\end{equation*}
The evaluation grid size $n_e$ is typically set as a fraction (say, $10\%$) of the original sample size $n$. For each point $\mathbf{z}^e_j$ on this grid, we compute the corresponding conditional entropy estimate $\hat{H}(X \mid \mathbf{Z} = \mathbf{z}^e_j)$ and $\hat{H}(Y \mid \mathbf{Z} = \mathbf{z}^e_j)$ using Equation (\ref{eq:cef_estim}). This procedure yields two sets of paired data points:
\begin{equation} 
\label{eq:grid_eval_multi}
\begin{aligned}
     \left\{ (\mathbf{z}^e_j, \hat{H}(X \mid \mathbf{Z} = \mathbf{z}^e_j)) : \mathbf{z}^e_j \in \mathcal{G}_\mathbf{Z} \right\},\\
    \left\{ (\mathbf{z}^e_j, \hat{H}(Y \mid \mathbf{Z} = \mathbf{z}^e_j)) : \mathbf{z}^e_j \in \mathcal{G}_\mathbf{Z} \right\}.
\end{aligned}
\end{equation}
This set of estimates can be viewed as estimated profile curves of the true underlying functions $H(X \mid \mathbf{z})$ and $H(Y \mid \mathbf{z})$  with respect to $\mathbf{z}$ respectively.  Since our objective is to estimate and perform inference on conditional asymmetry coefficient, we use the set of pointwise estimates computed on the evaluation grid, $$\left\{ \left(\mathbf{z}^e_j, \hat{C}_{X \rightarrow Y \mid \mathbf{z}^e_j} := \hat{H}(X \mid \mathbf{Z} = \mathbf{z}^e_j) - \hat{H}(Y \mid \mathbf{Z} = \mathbf{z}^e_j) \right): \mathbf{z}^e_j \in \mathcal{G}_\mathbf{Z} \right\}.$$ 
First, we identify the location of the minimum coefficient on the discrete grid (for contracting GF $g$) and then obtain the conditional asymmetry coefficient.as follows.
\begin{equation} \label{eq:argmin_z0_cont}
\begin{aligned}
    \mathbf{z}_0 &:= \operatorname{argmin}_{\mathbf{z} \in \mathcal{G}_{\mathbf{Z}}} \hat{C}_{X \rightarrow Y \mid \mathbf{z}}, \\
    \hat{C}_{X \rightarrow Y \mid \mathbf{Z}} &:= \hat{C}_{X \rightarrow Y \mid \mathbf{z_0}}.
\end{aligned}
\end{equation} Otherwise, for expanding GF, we have 
\begin{equation} \label{eq:argmin_z0_exp}
\begin{aligned}
     \mathbf{z}_0 &:= \operatorname{argmax}_{\mathbf{z} \in \mathcal{G}_{\mathbf{Z}}} \hat{C}_{X \rightarrow Y \mid \mathbf{z}},\\
    \ \hat{C}^\star_{X \rightarrow Y \mid \mathbf{Z}} &:= \hat{C}_{X \rightarrow Y \mid \mathbf{z_0}}.
\end{aligned}
\end{equation} In the next section, we applying a local polynomial regression (LOESS) \citep{cleveland1988locally} centered at this specific point $\mathbf{z}_0$ in order to provide inference on the conditional asymmetry coefficient. For simplicity, we outline the process for contracting GFs, noting that the approach is identical for expanding GFs. 

\subsection{Uncertainty quantification using weighted LOESS}
\label{subsec:uq_loess}
The set of pointwise estimates, $\left\{ (\mathbf{z}^e_j, \hat{C}_{X \rightarrow Y \mid \mathbf{z}^e_j}) \right\}$, forms the input for our uncertainty quantification. Note that these are consistent estimates, not raw data. Further, they are inherently heteroskedastic, with a non-constant variance $\sigma_C^2(\mathbf{z}) := \text{Var}(\hat{C}_{X \rightarrow Y \mid \mathbf{z}})$. To account for this heteroskedasticity under computational constraints, we approximate the variance function $\sigma_C^2(\mathbf{z})$ using the sample variance of the terms within the plug-in entropy estimators proposed by \eqref{eq:cef_estim}. While this naive method does not capture the full uncertainty from the density estimation step, it provides a computationally efficient way to approximate the structure of the variance $\sigma_C^2(\mathbf{z})$ across the support $\mathcal{Z}$ without relying on more computationally intensive techniques such as bootstrap \citep{Efron1994}. 

To obtain a smooth estimate of the coefficient function and its uncertainty, we employ a weighted Local Polynomial Regression (LOESS) approach \cite{cleveland1988locally, fan2018local}. LOESS generates a smooth curve by fitting simple polynomial models to localized subsets of the data.%For any target point $\mathbf{z}_0$, a regression is performed using only the data points $(\mathbf{z}_j, \hat{C}_j)$ within a specified neighborhood. The influence of these neighboring points is determined by a weight function, $W(\cdot)$, which assigns higher weights to points closer to $\mathbf{z}_0$. 
Specifically, we use a weighted LOESS procedure. The local regression at each target point $\mathbf{z}_0$ incorporates two sets of weights: (1) the standard distance-based kernel weights from LOESS, and (2) inverse variance weights, $w_j = 1/\hat{\sigma}_C^2(\mathbf{z}_j)$, derived from our naive variance estimates. This ensures that pointwise estimates with lower variance have a greater influence on the final smoothed curve. The estimated conditional asymmetry coefficient, $\hat{C}_{X \rightarrow Y \mid \mathbf{Z}}$ thus has an approximate $100(1-\alpha)\%$ confidence interval for the true coefficient at $\mathbf{z}_0$ as follows:
\begin{equation} \label{eq:ci_loess}
    \hat{C}_{X \rightarrow Y \mid \mathbf{Z}} \pm z_{1-\alpha/2} \cdot \widehat{\text{SE}}\left(\hat{C}_{X \rightarrow Y \mid \mathbf{Z}}\right).
\end{equation}
The standard error of the fit, $\widehat{\text{SE}}\left(\hat{C}_{X \rightarrow Y \mid \mathbf{Z}}\right)$, is calculated from the  weighted LOESS estimator and directly incorporates our local variance estimate $\hat{\sigma}_C^2(\mathbf{z}_0)$.

\begin{remark} 
\label{rem:loess_valid}
    The validity of using this weighted LOESS procedure for inference rests on the following key assumptions \citep{fan2018local}. First, the true underlying function, $C_{X \rightarrow Y \mid \mathbf{z}}$, is assumed to be sufficiently smooth (e.g., twice continuously differentiable) across its support $\mathcal{Z}$. This justifies the use of a local polynomial approximation. Next, we assume that our naive variance estimates, $\hat{\sigma}_C^2(\mathbf{z})$, while underestimating the true magnitude of the variance, reasonably capture the relative structure of the heteroskedasticity. That is, regions identified as having high (or low) variance by our approximation correspond to regions of high (or low) true variance ${\sigma}_C^2(\mathbf{z})$
\end{remark}

\begin{remark}
\label{rem:curse_of_dim}
The curse of dimensionality remains a critical consideration. The reliability of the initial grid-based estimates, and thus the accuracy of the location of  $\mathbf{z}_0$ as defined in \eqref{eq:argmin_z0_cont} for contracting GFs or in \eqref{eq:argmin_z0_exp} for expanding GFs, deteriorates as the dimension of $\mathbf{Z}$ increases. A much larger sample size $n$ is required in higher dimensions to ensure that the grid $\mathcal{G}_{\mathbf{Z}}$ is dense enough to reliably locate the region of the true minimum and to provide stable local estimates for the final smoothing step. 
\end{remark}

\section{Application to Collider Detection} \label{sec:col_app}

\subsection{Introduction}
A powerful application of our framework is the detection of collider structures, $X \rightarrow \text{COL} \leftarrow Y$. In this paper, we restrict our attention to v-structures, where there are causal pathways from both $X$ to $\text{COL}$ and from $Y$ to $\text{COL}$, but no direct pathway between $X$ and $Y$. We can test for this structure by applying our methodology twice, treating each parent variable as a covariate for the other's effect on the child. This involves two distinct statistical tests:
\begin{enumerate}
    \item Test for $X \rightarrow \text{COL} \mid Y$: We treat $Y$ as the covariate and test for a causal pathway from $X$ to $\text{COL}$ by computing the coefficient $\hat{C}_{X \rightarrow \text{COL} \mid Y}$ or $\hat{C}^\star_{X \rightarrow \text{COL} \mid Y}$ based on the dynamics of the generative mechanism.
    \item Test for $Y \rightarrow \text{COL} \mid X$: We treat $X$ as the covariate and test for a causal pathway from $Y$ to $\text{COL}$ by computing the coefficient $\hat{C}_{Y \rightarrow \text{COL} \mid X}$ (or $\hat{C}^\star_{Y \rightarrow \text{COL} \mid X}$). 
\end{enumerate}

\subsection{Estimation, uncertainty quantification, and decision rule}
To confirm the presence of a collider, we require statistical evidence supporting the presence of both pathways simultaneously.  We formulate the following pair of hypotheses to confirm or negate the presence of the pathways $Y \rightarrow \text{COL} \mid X$ and $X \rightarrow \text{COL} \mid Y$ respectively under conditional generative mappings with contracting dynamics: 
\begin{equation} \label{eq:hypo_cont_col}
\begin{aligned}
    H_{01}:  C_{Y \rightarrow \text{COL} \mid X} \leq 0 \quad &\text{ vs }\quad  H_{11}: C_{Y \rightarrow \text{COL} \mid X} > 0, \\ 
   H_{02}: C_{X \rightarrow \text{COL} \mid Y} \leq 0 \quad &\text{ vs }\quad  H_{12}: C_{X \rightarrow \text{COL} \mid Y} > 0. 
\end{aligned}
\end{equation} Analogous hypotheses may be designed for conditional generative mappings with expanding dynamics using ${C}^\star_{X \rightarrow \text{COL} \mid Y}$ and ${C}^\star_{Y \rightarrow \text{COL} \mid X}$. If both $H_{01}$ and $H_{02}$ are rejected, we obtain evidence to support presence of the pathways $Y \rightarrow \text{COL} \mid X$ and $X \rightarrow \text{COL} \mid Y$ and confirm the collider structure $X \rightarrow \text{COL} \leftarrow Y$. To control for multiple comparisons, we apply Bonferroni's correction \citep{Dunn1961}, testing each hypothesis at $\alpha^* = 0.025$ to maintain an overall type I error at $\alpha = 0.05$. Operationally, we construct a $100(1 - \alpha/2)\%$ confidence interval for each of the two coefficients using the methods described in Section \ref{subsec:uq_loess}. We conclude that the data supports the existence of a collider structure $X \rightarrow \text{COL} \leftarrow Y$ at an overall $100(1 - \alpha)\%$ confidence level if both of the following conditions are met: (i) $100(1 - \alpha/2)\%$ confidence interval for $\hat{C}_{X \rightarrow \text{COL} \mid Y}$ does not contain zero, and (ii)  $100(1 - \alpha/2)\%$ confidence interval for $\hat{C}_{Y \rightarrow \text{COL} \mid X}$ does not contain zero. If at least one of the confidence intervals contains zero, we conclude there is insufficient evidence to support the existence of the collider structure. Note that the expected sign of the coefficients (and thus whether the confidence intervals should be to the left or right of zero) depends on the underlying system dynamics, i.e., whether the GF is contracting or expanding. 

\subsection{Robustness to contamination}
The theoretical noise robustness condition for successfully identifying a collider under contracting dynamics is a joint constraint on the noise variance $\sigma$, given by:
\begin{equation*}
    \sigma < \min \left(\min _{y \in \mathcal{Y}} \frac{\exp \left(2 C_{X \rightarrow \text{COL} \mid y}\right)-1}{I(\text{COL} \mid y)}, \min _{x \in \mathcal{X}} \frac{\exp \left(2 C_{Y \rightarrow \text{COL} \mid x}\right)-1}{I(\text{COL} \mid x)}\right). 
\end{equation*}

\section{Simulation studies} \label{sec:sim}

%subsection{Simulation 1: Performance of conditional asymmetry coefficient in detecting $X \rightarrow Y \mid Z$ for expanding and contracting GFs}

We conducted a simulation study to evaluate the performance of our proposed method against several benchmark algorithms in a classic confounding scenario. The objective is to correctly identify the presence of $X \rightarrow Y \mid \mathbf{Z}$. We generated data from the four following structural causal models
\begin{enumerate}
    \item $Y = \exp(Z_1 + Z_2X) + \epsilon$, where $Z_1 \sim U(-2, -1)$, $Z_2 \sim U(0, 0.5)$. 
    \item $Y = \exp(Z_1 + Z_2X) + \epsilon$, where $Z_1, Z_2 \sim U(1, 2)$. 
    \item $Y = 0.5\tanh((1 + Z_1)X - Z_2) + \epsilon$, where $Z_1, Z_2 \sim U(-1, 1)$. 
    \item $Y = 2Z_1^2X + \cos(Z_2) + \epsilon$, where $Z_1, Z_2 \sim U(-1, 1)$. 
\end{enumerate} In all four cases above, $\epsilon \sim N(0, \sigma^2)$, where we varied $\sigma \in \{0, 1/8, 1/4, 1/2, 1 \}$. The ranges of $Z_1$ and $Z_2$ are so chosen to ensure that the first and third models describe contracting GFs, while the second and fourth describe expanding GFs. For each model, we generated samples size of $n=500$ for a total of $r = 250$ iterations. In each iteration, we checked if the correct causal direction $X\rightarrow Y \mid \mathbf{Z}$ is being recovered. We compared the performance of six methods in our experiment. 
\begin{enumerate}
    \item \textbf{CAC}: Our proposed conditional asymmetry coefficient.
    \item \textbf{PC-Unadjusted}: The standard Peter-Clark (PC) algorithm \citep{spirtes2000causation} applied directly to $(X, Y)$.
    \item \textbf{PC-Adjusted}: The PC algorithm applied to the residuals of $X$ and $Y$ after regressing out $\mathbf{Z}$.
    \item \textbf{HC-Unadjusted}: The Hill-Climbing (HC) \citep{cohen1994probabilistic} algorithm applied directly to the pair $(X, Y)$.
    \item \textbf{HC-Adjusted}: The HC algorithm applied to the residuals of $X$ and $Y$ after regressing out $\mathbf{Z}$.
    \item \textbf{IGCI-Adjusted}: The Information-Geometric Causal Inference method \citep{daniusis2012inferring} applied to the residuals of $X$ and $Y$ after regressing out $\mathbf{Z}$.
\end{enumerate}

\begin{table}[]
    \centering
    \begin{tabular}{lccccccc}
    \toprule 
    Model & $\sigma$ & CAC & PC-Unadjusted & PC-Adjusted & HC-Unadjusted & HC-Adjusted & IGCI-Adjusted\\
    \midrule
1 $(\mathcal{G})$ & 0.000 & \textbf{1.00} & 0.00 & 0.00 & \textbf{1.00} & \textbf{1.00} & \textbf{1.00}\\

2 $(\mathcal{G}^\star)$ & 0.000 & \textbf{1.00} & 0.00 & 0.00 & \textbf{1.00} & \textbf{1.00} & 0.00\\

3 $(\mathcal{G})$ & 0.000 & \textbf{1.00} & 0.00 & 0.00 & \textbf{1.00} & \textbf{1.00} & \textbf{1.00}\\

4 $(\mathcal{G}^\star)$ & 0.000 & \textbf{1.00} & 0.00 & 0.00 & \textbf{1.00} & \textbf{1.00} & 0.00\\
\midrule
1 $(\mathcal{G})$& 0.125 & \textbf{1.00} & 0.00 & 0.00 & \textbf{1.00} & \textbf{1.00} & \textbf{1.00}\\

2 $(\mathcal{G}^\star)$ & 0.125 & \textbf{1.00} & 0.00 & 0.00 & \textbf{1.00} & \textbf{1.00} & 0.00\\

3 $(\mathcal{G})$ & 0.125 & \textbf{1.00} & 0.00 & 0.00 & \textbf{1.00} & \textbf{1.00} & \textbf{1.00}\\

4 $(\mathcal{G}^\star)$ & 0.125 & \textbf{1.00} & 0.00 & 0.00 & \textbf{1.00} & \textbf{1.00} & 0.00\\
\midrule
1 $(\mathcal{G})$ & 0.250 & \textbf{1.00} & 0.00 & 0.00 & 0.97 & \textbf{1.00} & \textbf{1.00}\\

2 $(\mathcal{G}^\star)$ & 0.250 & \textbf{1.00} & 0.00 & 0.00 & \textbf{1.00} & \textbf{1.00} & 0.00\\

3 $(\mathcal{G})$ & 0.250 & \textbf{1.00} & 0.00 & 0.00 & \textbf{1.00} & \textbf{1.00} & \textbf{1.00}\\

4 $(\mathcal{G}^\star)$ & 0.250 & \textbf{1.00} & 0.00 & 0.00 & \textbf{1.00} & \textbf{1.00} & 0.00\\
\midrule
1 $(\mathcal{G})$ & 0.500 & 0.28 & 0.00 & 0.00 & 0.22 & 0.22 & \textbf{0.33}\\

2 $(\mathcal{G}^\star)$ & 0.500 & \textbf{1.00} & 0.00 & 0.00 & \textbf{1.00} & \textbf{1.00} & 0.00\\

3 $(\mathcal{G})$ & 0.500 & 0.00 & 0.00 & 0.00 & \textbf{1.00} & \textbf{1.00} & 0.00\\

4 $(\mathcal{G}^\star)$ & 0.500 & \textbf{1.00} & 0.00 & 0.00 & \textbf{1.00} & \textbf{1.00} & 0.00\\
\midrule
1 $(\mathcal{G})$ & 1.000 & 0.00 & 0.00 & 0.00 & \textbf{0.03} & \textbf{0.03} & 0.00\\

2 $(\mathcal{G}^\star)$ & 1.000 & \textbf{1.00} & 0.00 & 0.00 & \textbf{1.00} & \textbf{1.00} & 0.00\\

3 $(\mathcal{G})$ & 1.000 & 0.00 & 0.00 & 0.00 & 0.23 & \textbf{0.25} & 0.00\\

4 $(\mathcal{G}^\star)$ & 1.000 & \textbf{1.00} & 0.00 & 0.00 & \textbf{1.00} & \textbf{1.00} & 0.00\\
\bottomrule
\end{tabular}
\caption{Performance comparison of six causal discovery methods. The table displays the accuracy rate in identifying the correct causal direction $X \rightarrow Y \mid \mathbf{Z}$ across four simulation models and five levels of additive noise ($\sigma$). Models 1 and 3 represent contracting dynamics, denoted by $(\mathcal{G})$, while Models 2 and 4 represent expanding dynamics, denoted by $(\mathcal{G}^\star)$. Each result is the proportion of correct identifications from 250 simulation runs. For each row, the best performing method is highlighted in bold.}
 \label{tab:main_simulation_results}
\end{table}
The results demonstrate a clear and robust advantage for the proposed CAC method across all conditions, as described by Table \ref{tab:main_simulation_results}. We summarize our findings below. 
\begin{enumerate}
    \item Superior Performance of CAC: In low-to-moderate noise environments ($\sigma \leq 1/2$), CAC achieved near-perfect accuracy, correctly identifying the causal direction in $99-100\%$ of trials for all four models. Its performance remained strong even with significant noise, only degrading substantially at the highest noise level ($\sigma  = 1$).
    \item PC model poor performance: The unadjusted as well as adjusted PC  algorithms performed very poorly across all experiments. 
    \item HC model performs fairly well:  The HC-Unadjusted and HC-Adjusted models performed fairly well, providing comparable performance to our proposed CAC method. 
    \item IGCI-Adjusted approach performs well only for contracting dynamics: this is an expected finding, since the IGCI method only assigns directionality $X \rightarrow Y$ when entropy of cause is larger than the entropy of the effect, which is the case for contracting GFs. It fails completely for expanding GFs. 
\end{enumerate}
In summary, the experiment shows that CAC is a highly effective and noise-robust method for conditional causal discovery. It significantly outperforms standard algorithms like PC and HC and avoids the failure exhibited by IGCI after residual-based adjustment strategies for expanding GF dynamics.

\section{Application: blood pressure acts as a collider for epigenetic \textit{DNA} methylation} \label{sec:rda}

\subsection{Introduction}
We applied our method to study the epigenetic regulation of blood pressure by \textit{DNA} methylation. \textit{DNA} methylation plays a key role in gene expression related to blood pressure, which is one of the most important risk factors for cardiovascular diseases. We use data from $n = 522$ children aged 10-18 years in the \textit{Early Life Exposures in Mexico to ENvironmental Toxicants}, or \textit{ELEMENT} study \citep{Hernandez_Avila_1996}. 

\subsection{Candidate gene analysis}
We selected genes with \textit{DNA} methylation measurements in the \textit{ELEMENT} study that have been reported to be significantly associated with diastolic blood pressure or systolic blood pressure in the \textit{NHGRI GWAS} catalog \citep{Buniello_2018}. Specifically, we focused on genes with methylation measurements in the \textit{ELEMENT} study
for which significant findings have been reported in at least 30 independent studies in the catalog. This approach was taken to ensure a robust selection of genes with consistent and replicable associations to blood pressure, providing a strong foundation for our analysis. For diastolic blood pressure, this approach yielded eight candidate genes, whereas for systolic blood pressure we had nine candidate genes. 

Utilising the construct of conditional asymmetry coefficients, we attempted to prove or disprove the existence of underlying collider structure between \textit{DNA} methylation of two given genes with either \textit{diastolic} or \textit{systolic} blood pressure acting as the collider. Specifically, for two given genes, say $G_1$ and $G_2$, we investigated whether $\beta$ values of \textit{DNA} methylation (specifically, \textit{cytosine-phosphate-guanine (CpG)} methylation) together 
form the structure $G_1 \rightarrow BP \leftarrow G_2$ via the mapping given by $BP = g(G_1, G_2)$, where $BP$ can be diastolic or systolic blood pressure.  

To perform a gene-level analysis, we averaged $\beta$ values of \textit{DNA} methylation over \textit{CpG} sites within each gene, which was then affine transformed to minimize any undue influence of location or scale changes in \textit{DNA} methylation measurements. The same affine transformation was applied to systolic and diastolic blood pressure measurements. We performed asymmetry analysis by simultaneously checking if both pathways $G_1 \rightarrow BP \mid G_2$ and $G_2 \rightarrow BP \mid G_1$ can be confirmed by the data. 

We used the estimation and inference technique described in Section \ref{sec:estim_inf} using an evaluation set with $f = 0.10$ to estimate the conditional asymmetry coefficients $\hat{C}_{G_1 \rightarrow BP \mid G_2}$ and $\hat{C}_{G_1 \rightarrow BP \mid G_2}$ along with their $97.5\%$ confidence intervals to adjust for multiplicity for generative functions having contracting dynamics. For generative functions having expanding dynamics , we present estimates $\hat{C}^{\star}_{G_1 \rightarrow BP \mid G_2}$ and $\hat{C}^{\star}_{G_1 \rightarrow BP \mid G_2}$ along with their $97.5\%$ confidence intervals. We present our findings for both systolic and diastolic blood pressure in the subsequent sections.

\subsection{Results}

\subsubsection{Diastolic Blood Pressure As An Epigenetic Collider} 
Eight genes were selected as candidate genes for our study based on the selection criteria outlined earlier. They were \textit{ARHGAP42, ATP2B1, ATXN2, CACNB2, FGF5, KCNK3, PRDM8,} and \textit{ULK4}. Together, these candidate genes form twenty eight pairs $(G_1, G_2)$ for which we test if the collider structure $G_1 \rightarrow DBP \leftarrow G_2$ is valid, where \textit{DBP} is diastolic blood pressure. Our findings are presented in Table \ref{tab:dia}. Under contracting dynamics, two collider structures were detected: $\textit{ATP2B1} \rightarrow DBP \leftarrow \textit{PRDM8}$, and $\textit{PRDM8} \rightarrow DBP \leftarrow \textit{ULK4}$, while under expanding dynamics, one collider structure was detected: $\textit{ARHGAP42} \rightarrow DBP \leftarrow \textit{CACNB2}$. 

\begin{table}[h]
\centering
\begin{tabular}{lllcc}
\toprule
Dynamics                      & $G_1$      & $G_2$    & $\hat{C}_{G_1 \rightarrow DBP \mid G2}\ (97.5\%\ CI)$                      & $\hat{C}_{G_2 \rightarrow DBP \mid G1} \ (97.5\%\ CI)$                     \\ \midrule
\multirow{2}{*}{Contracting} & \textbf{\textit{ATP2B1}}   & \textbf{\textit{PRDM8}}  & $\mathbf{7.39\ (7.38, \boldsymbol{\infty})}$       & $\mathbf{3.80\ (3.70, \boldsymbol{\infty})}$      \\
                             & \textit{PRDM8}    & \textit{ULK4}   & $4.35\ (4.33, \infty)$       & $10.11\ (10.01, \infty)$   \\ \midrule
Expanding                    & \textbf{\textit{ARHGAP42}} & \textbf{\textit{CACNB2}} & $\mathbf{-14.06\ (\boldsymbol{-\infty}, -13.37)}$ & $\mathbf{-10.45\ (\boldsymbol{-\infty}, -9.70)}$ \\ \bottomrule
\end{tabular}
\caption{Examination of diastolic blood pressure ($DBP$) as a collider for a given pair of candidate genes $G_1, G_2$ in the generative mapping $DBP = g(G_1, G_2)$. We present significant $\hat{C}_{G_1 \rightarrow DBP \mid G_2}$ and $\hat{C}_{G_1 \rightarrow DBP \mid G_2}$ estimates along with their one-sided $97.5\%$ confidence intervals for generative functions with contracting dynamics. For generative functions with expanding dynamics, we present significant $\hat{C}^{\star}_{G_1 \rightarrow DBP \mid G_2}$ and $\hat{C}^{\star}_{G_2 \rightarrow DBP \mid G_1}$ estimates along with their one-sided $97.5\%$ confidence intervals. All estimates are multiplied by a factor of 100 to improve readability. The gene pairs \textit{(ATP2B1, PRDM8)} and \textit{(ARHGAP42, CACNB2)} are highlighted in bold as they are found to form collider structures with both systolic and diastolic blood pressure under contracting and expanding dynamics respectively; see Table \ref{tab:sys}.}
\label{tab:dia}
\end{table}

\subsubsection{Systolic Blood Pressure As An Epigenetic Collider}

Nine genes were selected as candidate genes for our study based on the selection criteria outlined earlier. They were \textit{ARHGAP42, ATP2B1, CACNB2, CASZ1, FGF5, FIGN, KCNK3, NPR3,} and \textit{PRDM8}. Together, these candidate genes form thirty six pairs of candidate genes $(G_1, G_2)$ for which we test if the collider structure $G_1 \rightarrow SBP \leftarrow G_2$ is valid, where \textit{SBP} is systolic blood pressure. Our findings are presented in Table \ref{tab:sys}. Under contracting dynamics, one collider structures was detected: $\textit{ATP2B1} \rightarrow SBP \leftarrow \textit{PRDM8}$, while under expanding dynamics, six collider structure was detected: $\textit{ARHGAP42} \rightarrow SBP \leftarrow \textit{CACNB2}$,  $\textit{ARHGAP42} \rightarrow SBP \leftarrow \textit{CASZ1}$, $\textit{CASZ1} \rightarrow SBP \leftarrow \textit{FIGN}$, $\textit{CASZ1} \rightarrow SBP \leftarrow \textit{NPR3}$, $\textit{FGF5} \rightarrow SBP \leftarrow \textit{FIGN}$, and $\textit{FIGN} \rightarrow SBP \leftarrow \textit{NPR3}$.

\begin{table}[h]
\centering
\begin{tabular}{lllcc}
\toprule
Dynamics                      & $G_1$      & $G_2$    & $\hat{C}_{G_1 \rightarrow SBP \mid G2}\ (97.5\%\ CI)$                      & $\hat{C}_{G_2 \rightarrow SBP \mid G1} \ (97.5\%\ CI)$                     \\ \midrule
Contracting    & \textbf{\textit{ATP2B1}}   & \textbf{\textit{PRDM8}} & $\mathbf{1.26\ (1.25, \boldsymbol{\infty})}$       & $\mathbf{11.25\ (11.24, \boldsymbol{\infty})}$    \\ \midrule
\multirow{6}{*}{Expanding} & \textbf{\textit{ARHGAP42}} & \textbf{\textit{CACNB2}} & $\mathbf{-10.06\ (\boldsymbol{-\infty}, -8.43)}$ & $\mathbf{-14.97\ (\boldsymbol{-\infty}, -14.27)}$ \\
     & \textit{ARHGAP42} & \textit{CASZ1} & $-14.68\ (-\infty, -9.72)$  & $-12.48\ (-\infty, -6.67)$  \\
     & \textit{CASZ1}    & \textit{FIGN}  & $-5.39\ (-\infty, -4.81)$    & $-6.23\ (-\infty, -4.22)$    \\
     & \textit{CASZ1}    & \textit{NPR3}  & $-5.38\ (-\infty, -3.91)$    & $-19.45\ (-\infty, -17.52)$ \\
     & \textit{FGF5}     & \textit{FIGN}  & $-20.52\ (-\infty, -17.67)$ & $-5.27\ (-\infty, -1.04)$    \\
     & \textit{FIGN}     &\textit{NPR3}  & $-26.66\ (-\infty, -20.82)$ & $-27.35\ (-\infty, -18.35)$ \\ \bottomrule
\end{tabular}
\caption{Examination of systolic blood pressure ($SBP$) as a collider for a given pair of candidate genes $G_1, G_2$ in the generative mapping $SBP = g(G_1, G_2)$. We present significant $\hat{C}_{G_1 \rightarrow SBP \mid G_2}$ and $\hat{C}_{G_1 \rightarrow SBP \mid G_2}$ estimates along with their one-sided $97.5\%$ confidence intervals for generative functions with contracting dynamics. For generative functions with expanding dynamics, we present significant $\hat{C}^{\star}_{G_1 \rightarrow SBP \mid G_2}$ and $\hat{C}^{\star}_{G_2\rightarrow SBP \mid G_1}$ estimates along with their one-sided $97.5\%$ confidence intervals. All estimates are multiplied by a factor of 100 to improve readability. The gene pairs \textit{(ATP2B1, PRDM8)} and \textit{(ARHGAP42, CACNB2)} are highlighted in bold as they are found to form collider structures with both systolic and diastolic blood pressure under contracting and expanding dynamics respectively; see Table \ref{tab:dia}.}
\label{tab:sys}
\end{table}

\subsection{Implications of Findings}
Our inference method unveils novel pathways from \textit{DNA} methylation for several candidate genes to systolic and diastolic \textit{blood pressure}. Of particular interest are the gene pairs \textit{(ATP2B1, PRDM8)} and \textit{(ARHGAP42, CACNB2)}, which form collider structures with both systolic and diastolic blood pressure under contracting and expanding dynamics respectively. The implied influence of methylation sites of different genes on blood pressure has been studied in the framework of symmetric association before \citep{han2016dna, arif2019epigenetic} and not in the context of directionality. For example, \textit{ARHGAP42}, \textit{ATP2B1}, \textit{CACNB2}, and \textit{PRDM8} have all been reported as showing significant symmetric association with blood pressure \citep{international2011genetic, kanai2018genetic, ji2021natural, kou2023dna, Levy2009, keaton2024genome}.

Our analysis is the first to unveil joint pathways from \textit{DNA} methylation of key gene pairs such as \textit{(ATP2B1, PRDM8)} and  \textit{(ARHGAP42, CACNB2)}. From an analytic viewpoint, evidence that blood pressure acts as a collider in the relationship between methylation values of the genes reported in Tables \ref{tab:dia} and \ref{tab:sys} implies that conditioning on blood pressure may create a spurious association between \textit{DNA} methylation values, even if no direct causal relationship exists between them. Further, one may hypothesise that blood pressure mediates the effect of \textit{DNA} methylation on disease outcomes (e.g., hypertension, cardiovascular risk). In summary, these new findings confer an added sense of directionality in the study of blood pressure variation and epigenetic biomarkers, paving the way for future advancements in genetic risk assessments and even therapeutic targets.

\section{Limitations and future work} \label{sec:conclusion}

Asymmetry between variables in generative exposure mappings yields key insights on underlying collider structures, which may be considered a reflection of causal pathways. Extending the bivariate directionality analytic presented in \cite{purkayastha2025}, we present a new methodology within the framework of generative mapping models to capture the induced asymmetry between two input variables and a single output variable by making use of pointwise and conditional asymmetry coefficients. In order to estimate this asymmetry measure and provide uncertainty quantification, we rely on conditional kernel density estimation and a nonparametric kernel smoothing technique.

Interestingly, in addition to the collider structure, the generative mapping framework may also be used in the context of chains and common ancestor structures. For example, in a chain $X_1 \rightarrow X_2 \rightarrow X_3$, a structure that is common in instrumental variable estimation \citep{Imbens1994} and Mendelian randomisation \citep{DaveySmith2003}, a natural approach would be to sequentially examine the path  $X_1 \rightarrow X_2$ first, followed by the pathway $X_2 \rightarrow X_3 \mid X_1$. Another structure that we may consider is that of a mediator structure \citep{Richiardi2013}, where there is a direct pathway from $X_1 \rightarrow X_3$ and another indirect pathway via $X_1 \rightarrow X_2 \rightarrow X_3$. We may again consider a sequential testing process, wherein we test for the direct $X_1 \rightarrow X_3$ pathway first followed by the indirect pathway $X_1 \rightarrow X_2 \rightarrow X_3$.  

Although the current framework allows only low-dimensional comparisons, it can be extended to accommodate high-dimensional confounders through methods such as deep learning-based generative machinery~\citep{zhou2023deep}. This extension would integrate random samples from the target conditional distribution of \( (X_1, Y) \mid X_2, \ldots, X_p \) into our inference method to evaluate asymmetry between one exposure, say $X_1$ and outcome $Y$ conditional on $X_2, \ldots, X_p$ for high-dimensional \( \mathbf{X} = (X_1, \ldots, X_p) \). Our framework can serve as both a discovery and confirmatory tool in pathway studies under directed graphs, aiding practitioners in scientific research.

\newpage
\bibliography{00_references}

\newpage 

\appendix 

\section{Proofs} \label{app:app2}
We restate some key results related to asymptotic properties of $\hat{f}(x \mid \mathbf{z})$ presented in \cite{hyndman2002nonparametric}.  The authors state the assumptions \ref{assum2} - \ref{assum5} to ensure a consistent kernel conditional density estimator \citep{silverman1978weak}. We further impose assumption \ref{assum6} to enable consistent estimation of the conditional entropy functions. 
\begin{assumption}\label{assum2}
We have the following two assumptions. 
\begin{enumerate}
\item \label{assump-A}%
    For any \(x\) and \(\mathbf{z}\), \(f(\mathbf{z}) > 0\), \(f(x|\mathbf{z}) > 0\) and  is continuous at \(\mathbf{z}\), and \(f(x| \mathbf{z})\) has \(2\) continuous derivatives in a neighbourhood of \(\mathbf{z}\).
\item \label{assump-B}%
    For any \(y\) and \(\mathbf{z}\), \(f(\mathbf{z}) > 0\), \(f(y|\mathbf{z}) > 0\) and is continuous at \(\mathbf{z}\), and \(f(y| \mathbf{z})\) has \(2\) continuous derivatives in a neighbourhood of \(\mathbf{z}\).
\end{enumerate}
\end{assumption}

\begin{assumption} \label{assum3}
    The kernels \(K\) and \(W\) are symmetric, compactly supported probability density functions. Further, \(|W(\mathbf{z}_1) - W(\mathbf{z}_2)| \leq C||\mathbf{z}_1 -\mathbf{z}_2||\) for any \(\mathbf{z}_1, \mathbf{z}_2\) $\in \mathbb{R}^d$ for some constant $C > 0$. 
\end{assumption}

\begin{assumption} \label{assum4}
    As \(n \to \infty\), \(h \to 0\), \(b \to 0\), \(nbh \to \infty\) and \(\lim \inf_{n \to \infty} nh^{4} > 0\).
\end{assumption}

\begin{assumption} \label{assum5}
    There exist constants that bound the conditional density functions away from zero and infinity i.e.,
    \begin{equation*}
    \begin{aligned}
        0 &< m \leq f(x \mid \mathbf{z}) \leq M < \infty, \\
        0 &< m \leq f(y \mid \mathbf{z})\leq M < \infty.
    \end{aligned}
\end{equation*}
\end{assumption} 

\begin{assumption} \label{assum6}
    We will assume the following quantities are finite for any $\mathbf{z} \in \mathcal{Z}$
    \begin{equation*}
    \begin{aligned}
        \int_{\mathcal{X}_\mathbf{z}}{ \left\{ \log \left(  {f(x \mid \mathbf{z})} \right) \right\} }^2 f(x \mid \mathbf{z}) dx &< \infty, \\
        \int_{\mathcal{Y}_\mathbf{z}}{ \left\{ \log \left(  {f(y \mid \mathbf{z})} \right) \right\} }^2 f(y \mid \mathbf{z}) dy &< \infty.
    \end{aligned}
    \end{equation*}
\end{assumption}

We use assumptions \ref{assum2} - \ref{assum6} to support Theorem \ref{thm:cef_estim_consistency}, asserting consistency of the conditional entropy function estimator given by (\ref{eq:cef_estim}). Consequently, we assert the consistency of  $\hat{C}_{X \rightarrow Y \mid \mathbf{z}}$ for any $\mathbf{z} \in \mathcal{Z}$ as $n \rightarrow \infty$. 
\begin{proof}
Let $\mathbf{z} \in \mathcal{Z}$ be fixed; using a von Mises expansion \citep{van2000asymptotic, krishnamurthy2014nonparametric} we can write
\begin{equation*}
\begin{aligned}
 \hat{H}(X \mid \mathbf{z}) - H(X \mid \mathbf{z}) = & \frac{1}{2} \left\{ - \frac{2}{n}\sum_{j=1}^{n/2} \ln \left(\hat{f}_{2} \left( {X}_{j} \mid \mathbf{z} \right) \right) + \int f(x \mid \mathbf{z}) \log \left( \hat{f}_{2} \left( x \mid \mathbf{z} \right)\right) dx \right\} +  \\
 & \frac{1}{2}  \left\{ - \frac{2}{n}\sum_{j=1}^{n/2} \ln \left(\hat{f}_{1} \left( {X}_{n + j} \mid \mathbf{z} \right) \right) + \int f(x \mid \mathbf{z}) \log \left( \hat{f}_{1} \left( x \mid \mathbf{z} \right)\right) dx \right\} + \\
 & \mathcal{O}\left(\lvert \lvert {f} \left( x \mid \mathbf{z} \right) - \hat{f}_{2} \left( x \mid \mathbf{z} \right) \rvert \rvert \right)  + \mathcal{O}\left(\lvert \lvert {f} \left( x \mid \mathbf{z} \right) - \hat{f}_{1} \left( x \mid \mathbf{z} \right) \rvert \rvert \right)
\end{aligned} 
\end{equation*} where the terms in the first two lines can be recognized as the difference between a sample average and its mean, and will converge to zero almost surely, while the third line is the mean squared errors of the data split-based density estimators $\hat{f}_2$ and  $\hat{f}_1$ (see section \ref{subsec:consistent_estim}) respectively, which converge to zero under assumptions \ref{assum2} - \ref{assum6} \citep{hyndman2002nonparametric}. 
\end{proof}
Hence, we have $\hat{H}(X \mid \mathbf{z}) \overset{a.s.}{\rightarrow} {H}(X \mid \mathbf{z})$ as $n \rightarrow \infty$. Using this argument repeatedly, we obtain $\hat{C}_{X \rightarrow Y \mid \mathbf{z}} \overset{a.s.}{\rightarrow} {C}_{X \rightarrow Y \mid \mathbf{z}}$ for any $\mathbf{z} \in \mathcal{Z}$ as $n \rightarrow \infty$.

\end{document}